\documentclass[prb]{revtex4}
\usepackage{epsf}
\begin{document}
\title{General Relativity for Pedestrians - First 6 lectures}
\author{Patrick Das Gupta}
\affiliation{Department of Physics and Astrophysics, University of Delhi, Delhi - 110 007 (India)}
\email{pdasgupta@physics.du.ac.in}
%\documentclass[11pt]{article}        
%\voffset=-1.0cm
%\hoffset=-1cm
%\textwidth=6in
%\textheight=9in
%\renewcommand{\baselinestretch}{1.3}
%\def\dis{\displaystyle}
%\def\barr{\begin{array}}
%\def\earr{\end{array}}
%\begin{document}                   
%\begin{center}
%{\large \bf Patrick Das Gupta}\\[2ex]
%{\large \bf Department of Physics and Astrophysics}\\

%{\large \bf University of Delhi} \\[2ex]
\begin{abstract}
 The 2017 Nobel Prize in physics awarded to Rainer Weiss, Barry C. Barish and Kip S. Thorne has generated unprecedented interest in gravitational waves (GWs). These notes are based on my lectures on various occasions - in the University of Delhi as well as in different GW schools held in India following the exciting direct detection of GWs. I discuss GW flux and luminosity while pointing out a curious aspect associated with the latter -   physical dimensions of $c^5/G$ as well as third time derivative of mass quadrupole moment are that of luminosity. Formation of primordial black holes in the early universe and progenitors of fast radio bursts could have generated  GW luminosity comparable to the Planck luminosity, $c^5/G$. I also address the issue of black hole thermodynamics in connection with the  GW150914 event,  demonstrating that this event is consistent with Hawking's black hole area theorem. In the last section, as an illustrative exercise, I estimate the GW amplitude expected from the fast moving plasma bullets that have been shot out from the vicinity of the carbon star V Hydrae, as reported recently  by Sahai et al. (2016). 
\end{abstract}
\maketitle
\section{Introduction}
Gravity is universal. Everything creates gravity as well as gets affected by gravity since everything has  mass (or, equivalently, energy). However, unless the mass of an object is very large, the gravity it generates  is very weak. 

 According to Newton's laws of gravity,  acceleration of a test particle  due to the gravitational field of a massive object, is proportional to latter's mass, is directed towards the massive body, is inversely proportional to the square of the distance between the two objects, and furthermore,  is independent of the test particle's  mass. 

Newton, with a flash of brilliance, had  realized  that  Moon's  orbiting of  the Earth is nothing but  its continuous fall under Earth's gravity. He estimated from the Moon's orbital period of about 28 days that its acceleration directed towards us due to Earth's  gravity is smaller than the acceleration of an apple falling on Newton's head by a factor of square of the ratio of Earth's radius  to the Moon-Earth distance. With a leap of generalization,  Newton  deduced the inverse square law for gravity.

But Newton's theory is  inconsistent with special theory of relativity (STR). If the Sun were to disappear at this instant then Newton's theory predicts that  at this very instant the Earth will  fly off tangentially (in the absence of a centripetal force). But according to STR, no information can travel faster than the speed of light, so the disappearance of Sun cannot instantaneously affect   Earth's trajectory.

Einstein corrected the situation by proposing in 1915, about hundred years back,  a  consistent  theory of gravity  through his theory of general relativity. General theory of relativity (GTR) is a relativistic theory of gravitation. GTR is  based on the observation that the trajectory of a test particle in any arbitrary gravitational field is independent of its inertial mass $ m$ (as  the acceleration does not dependent on $m$), and therefore, it must be the geometry of  space-time that determines test particle trajectories.

 Note that for no other force, acceleration of a test particle is independent of its inertial mass (e.g. in the presence of electromagnetic fields, acceleration of a test charge  is proportional to the ratio of its charge to mass).

\section{Gravity, Inertial Frames and Equivalence Principle}
In an inertial frame, according to Newtonian laws of gravity, the magnitude of gravitational force between two objects 1 and 2,   separated by a distance $d$, is given by $ F= - \frac {GM_1 M_2} {d^2}$, where $M_1$ and $M_2$  are the gravitational masses. Gravitational mass $ M $ plays the role of gravitational charge. This is analogous to  the Coulombic case of  electric force  between electric charges.  
But the magnitude of acceleration of object 1 due to this gravitational force is  $ a= \frac {F} {m_1} $  where $m_1$  is the inertial mass (which appears in $ F=m a $ or momentum $p = m v $, etc.).

 But from experiments we know that $ a $ is independent of $ m_1 $ (e.g. Galileo's,  and later the torsion balance experiments). In other words, the gravitational acceleration of a test particle is independent of its inertial mass. This is called the weak equivalence principle. Because of this, the gravitational mass $M_1$  divided by the  inertial mass $ m_1$  has to be a constant for all objects. Hence, we can choose units for masses so that this constant has the value 1.

But what is an inertial frame? One operational way of defining an inertial frame is that it is a frame of reference in which if there is no real force acting on an object, then the object either remains at rest or it moves with a uniform velocity. Such a definition rules out an accelerating frame to be an inertial frame.

But the key condition is that there should be no real force acting on the object. One can always shield it from electromagnetic forces, and weak and nuclear forces are anyway short-ranged. But what about gravity? Anything that has energy has mass too, and therefore will be a source of Newtonian gravity. So, how does one create a frame that has  no gravity in it?

Einstein, with his brilliant insight,  offered   an ingenious solution to this predicament. Basically, he  employed the Galilean-Newtonian weak equivalence principle which states that in a given region, acceleration of a body due to external gravity is independent of its inertial mass. Imagine that a small bundle of test particles are freely falling in an arbitrary gravitational field. Since their accelerations due to gravity are nearly identical as their relative separations are small, if one were to sit on one such particle and observe the rest, one would find that the other test particles are freely floating as though gravity has simply disappeared!

This is  Einstein's  principle of equivalence according to which no matter how strong or how time varying the gravity is, one can always choose a small enough frame of reference, for a sufficiently small time interval such that gravity vanishes in this frame. So, one has obtained a truly inertial frame, albeit of  a limited size one! In other words, from Einstein's argument, no matter where, one can always construct a local inertial frame, the size of the frame depending on the scale on which the gravity varies.
However, we have a queer situation here: according to a freely falling observer A, there is no gravitational force in her/his neighbourhood, while on the other hand, according to an outside  observer B who is at rest on the surface of the earth, there is gravity acting on the falling observer.  In the Newtonian paradigm, the existence of a genuine force cannot depend on the choice of  frames of reference.

 To highlight  the above point  further, let us look at  Einstein's equivalence principle  from another angle. Consider a frame of reference that is far removed from sources of gravitation so that there is no external gravity felt by an observer C  anywhere in this frame. But if this frame C  is accelerating with respect to an inertial frame (i.e. C  is a non-inertial frame), then the observer C will experience a pseudo-gravitational force. No measurement in C can distinguish between real gravity and the pseudo-gravity if weak equivalence principle is correct.  Is gravity then a `real force'?  We will see shortly how this perplexing issue  is resolved in Einstein's GTR.

\section{Gravity and Special Theory of Relativity} 
Newton's theory of gravitation also demands that the gravitational force be instantaneously transmitted by the source to the test particle, since it is inversely proportional to the square of the  instantaneous separation between the two.  Instant transmission is unsatisfactory, as Einstein's special theory of relativity demands that no physical effect can  propagate  faster than $c= 3\ \times 10^8 m\ s^{-1}$, the speed of light in vacuum or,  for that matter, speed of  any particle with zero rest mass. This ensues from the relativistic expression for energy $E$ of a free particle with rest mass $m$ given by,
$$ E = \frac {m c^2} {\sqrt { 1-\frac {v^2} {c^2}}} \eqno(1)$$
From the above, it is evident that if $v > c $, the energy becomes pure imaginary, ruling out faster than light motions.
 Clearly, gravitational  theory needs  to  incorporate relativity. But, how? The clue comes from Einstein's version of equivalence principle. We have seen in the previous section that weak equivalence principle guarantees that in the presence of gravitation it is always possible to choose a limited size frame of reference for a short enough time in which gravity disappears (e.g. a freely  falling frame).  This limited  region  constitutes a local inertial frame of reference, so that a  Cartesian coordinate system  can be set up for specifying spatial coordinates here,  and clocks can be arranged to measure proper time. Such a coordinate system is referred to as the Minkowskian coordinate system.  
Therefore, in this local inertial frame (LIF), the laws of physics (other than the gravitational phenomena)  must take the same form as they do in special theory of relativity. The proper distance $ds $  between two nearby events  in the LIF with space-time coordinates $x^\mu$ = $(ct,x,y,z)$ and $x^\mu + dx^\mu$ = $(ct +cdt,x + dx, y+dy, z+dz)$   is  evaluated, as in STR, using,
$$ds^2 = c^2 dt^2 - dx^2 - dy^2 - dz^2 \equiv \eta_{\mu \nu}dx^\mu dx^\nu \ . \eqno(2)$$
Note that in eq.(2),  $x^i$, i=1,2,3 are the  Cartesian coordinates of the event, and $\eta_{\mu \nu} $ is the Minkowski metric with $\eta_{00}=1= -\eta_{ii}$, rest of the off-diagonal components of the metric being zero. Einstein summation convention has been used in eq.(2), so that repetition of Greek indices imply summation over 0,1,2 and 3.

{\bf {(From now on we will adopt the Einstein summation convention wherein whenever a Greek index repeats in an expression it means that the expression is being summed over with the index running from 0 to 3.)}}
 
 STR not only proclaimed that time is the fourth dimension but also  necessitated a departure from the  Euclidean notion of distances. For events that are not causally connected $ds^2$ is negative, which was unthinkable in  Euclidean paradigm.
  
 In other words, by going over to a small freely falling frame and choosing a locally  inertial or  Minkowskian coordinate system, one manages not only  to make gravity vanish locally but also express  the non-gravitational laws of physics exactly as one does in special relativity. But,  what about the laws pertaining to gravitation itself? And, what if one requires to express  laws of physics over larger regions of space-time? 
 
Let us first deal with the simplest and hypothetical `gravitational' set up  -  the case of uniform and static   gravity, so that  the acceleration vector due to gravity is  same everywhere, and at all times. But this situation, according to the equivalence principle is identical to   zero  gravity  case,  for,  one has to just consider a freely falling reference frame as large as and for as long as one wants, and in this frame  gravity simply vanishes. Hence, one can choose Minkowski coordinates globally and eq.(2) describes the space-time geometry everywhere in such a frame.  Thus, uniform gravity everywhere is equivalent to zero gravity. 

Our next case is: gravity around a massive, spherically symmetric body  of radius $R$ and mass $M$.  From the  Newtonian point of view, the acceleration due to gravity caused by it at any external point P,  is inversely proportional to  the distance between P and the centre of the massive object, and is directed towards the centre. Now, if one considers a large freely falling frame (LFFF)  at  an  initial distance $ d \gg R $, then does gravity totally vanish in this frame?

 Clearly the answer is no. For, if one takes two test particles 1 and 2 separated  by a vector  $\vec L $,  that are freely falling along with this LFFF,  then if   $\vec L $ is perpendicular to the radial direction of fall, an observer in LFFF will notice  1 and 2  to be accelerating towards each other with a magnitude,
$$ a_{12} \approx \frac {G\ M \ L} {d^3} \eqno(3) $$
because each of the particles will be accelerating radially towards the centre of the massive body. 

On the other hand, if $\vec L $ was along the radial direction of the free fall, the observer in LFFF would measure 1 and 2 to be accelerating away from each other with a magnitude given by eq.(3), as the particle nearer to the massive object would be falling with a greater acceleration than the one farther away. These are nothing but instances of tidal acceleration, ubiquitous whenever gravity is non-uniform. 

Although gravitational acceleration vanishes in a local inertial frame (LIF), tidal acceleration does not. It is just that in a LIF, the magnitude of  $\vec L $ is small as the frame itself is of  limited size, so that according to eq.(3), the value of the tidal acceleration  is negligibly small here. But when the frame is large, its different parts encounter varying degree of  tidal stretching or tidal compression. For instance, we do not experience Sun's gravity as  Earth is  freely  falling towards the Sun. 

 Nevertheless   oceans exhibit high and low tides, since our planet  is large enough for   Sun's  tidal forces to be non-negligible.
The above example shows that, in general, one cannot eliminate the effects of gravitation entirely. The LIFs, however, are very useful through the use of STR, for  the  extraction of physical meanings of various mathematical expressions.

 Since one cannot have in general global inertial (i.e. Minkowski) coordinates in the presence of gravity,  it is necessary to develop a formalism that employs arbitrary coordinates like curvilinear coordinates in the analysis.  One can motivate the necessity of using coordinate systems other than the Minkowskian ones,  from a physical standpoint.

 Consider the case of a sufficiently large reference frame that is made up  of 3-dimensional  Cartesian grid of standard rods with clocks  arranged at their intersections. Such a framework of Minkowskian coordinate system  cannot be maintained as a  LFFF, when there is  non-uniform gravitation present, because of the following reason. 
 
  From STR,  the condition that nothing can travel faster than $c$ implies that  no object can be  absolutely rigid.  Otherwise, one could simply  transfer energy (and therefore, signals) from one spatial point to another with infinite speed, just  by tapping one end of a long `rigid' rod causing the other end to move instantaneously. Now, non-uniform gravity would mean that different portions of the LFFF would fall with  different accelerations,  leading to stretching  and compression of the (initially) cubical grid of rods and clocks (forces other than gravity will enter the analysis), so that it is no longer possible to maintain a global Minkowskian coordinate system in any LFFF. Curvilinear coordinates, therefore,  become indispensable in relativistic gravitational physics.

In STR, square of the proper (i.e. Lorentz invariant) distance between any two infinitesimally events is given by eq.(2) when Minkowskian coordinates are chosen. Preceding arguments make it clear that when gravitation is included, one would need to modify eq.(2) and,  instead of the  Minkowski metric, one would  require a general metric tensor. Similarly,  the concept of tidal acceleration has to be made precise from the point of view of arbitrary frames of reference that use general coordinate systems.

\section{Curvilinear coordinates, scalar, vector and tensor fields}
Let an {\bf event} occur at some space-time point P, which is assigned a coordinate $x^\mu (\mbox{P}) $ by an  observer O, with $\mu=0,1,2,3$, which are four real numbers corresponding to one time and three space coordinates. The same point P, in general, will have a different coordinate  $x^{\prime\ \mu } (\mbox{P}) $ according to another observer O'. Note that the observers O and O' may not be inertial observers so that the coordinates  $x^\mu  $ and $x^{\prime\ \mu }  $ are, in general, curvilinear coordinates. A {\bf space-time manifold}  is defined to be the set of all events. In GTR, the mathematical forms of physical laws remain the same even when one makes an arbitrary coordinate transformation.

In general, any  arbitrary event belonging to a space-time manifold can be assigned  coordinates $x^\mu $ and $x^{\prime \mu}  $ by O and O', respectively. Since labeling  of an event with coordinates  by an observer  involves a mapping from the space-time manifold to $R^4$,   it follows that there is a mapping between $x^\mu $ and $x^{\prime \mu}  $, i.e. there exists a  function that relates coordinate system employed by O  to that of O'. Therefore, one may either treat  $x^{\prime \mu}  $ to be a continuous and differentiable function of $x^\alpha $  or  vice-versa (i.e. $x^\mu $ as a smooth function of  $x^{\prime \alpha}  $), with $\mu \ \ , \alpha =0,1,2,3$. 

The demand for a smooth function can be justified from a classical physics standpoint that  events can be arbitrary close to each other  with no holes or discreteness in the space-time manifold.  Going from one set of coordinates $x^\alpha $ to another set $x^{\prime \alpha}  $ is called a {\bf general coordinate transformation}. In a given coordinate system, each coordinate component of an event is functionally independent of the other coordinate so that,
$$\frac {\partial x^\mu} {\partial x^\nu} = \delta ^\mu _\nu \eqno(4)$$ 
Consider  some  physical variable (e.g. comoving energy density or pressure of a fluid) that can be described by   observer O as a real valued function $ \phi (x^\alpha) $ of space-time coordinates. Observer O', however, will  find the same physical variable  to be represented by a different function $ \phi^\prime  (x^{\prime \alpha}) $. The function $\phi $ is said to be  a {\bf scalar field}  if  everywhere on the space-time manifold,
$$ \phi^\prime  (x^{\prime \alpha}) = \phi (x^\alpha ) $$
given that $x^\alpha $ and $x^{\prime \alpha} $ are the space-time coordinates assigned by observers O and O', respectively, to the same event. 

 Physically, what a scalar field signifies is that,  at every event  P, the value of the physical variable   $ \phi (x^\alpha (P)) $  as measured by the observer O  is identical to the value   $ \phi^\prime  (x^{\prime \alpha} (P))  $ as measured by O', although  the functional form of the physical variable depends on the observer. 

 How does one define vector components when one is using  general curvilinear coordinates? Intuitively, a vector has magnitude as well as direction, and hence  it resembles an arrow. Suppose, we have two events P and P' which are  temporally as well as spatially near each other, so that they have coordinates $ x^\mu $ and $x^\mu + dx^\mu $, respectively, for  observer O. Clearly, the directed line PP' from P to P' looks like an infinitesimally short arrow and thereby  qualifies to be called a vector with $dx^\mu $ as the vector components. 
 
  According to O', however, the directed line PP' has components $dx^{\prime \mu} $ since P and P' have coordinates $x^{\prime \mu} $ and $x^{\prime \mu} + dx^{\prime \mu} $, respectively, in her/his frame. The relation between the components is given by the usual rules of partial derivatives.      
  
  $$dx^{\prime \mu} = \frac {\partial x^{\prime\mu}} {\partial x^\nu} dx^ \nu \eqno(5)$$
The above equation suggests that a contravariant vector field $V^\mu (x^\alpha)$ ought to be defined as an entity that transforms under a general coordinate transformation   $x^\gamma \rightarrow x^{\prime \gamma}$ in the following manner,
$$V^\mu (x^\alpha) \rightarrow V^{\prime \mu} (x^{\prime \alpha}) = \frac {\partial x^{\prime\mu}} {\partial x^\nu} V^\nu (x^\alpha)\eqno(6)$$

The next question that arises is: What about objects like $\frac {\partial \phi (x)} {\partial x^\mu} $, where $\phi (x^\alpha)$ is a scalar field? Let us see how this entity transforms under general coordinate transformation. When $x^\gamma \rightarrow x^{\prime \gamma}$, we find that,
$$\frac {\partial \phi (x)} {\partial x^\mu}  \rightarrow \frac {\partial \phi^\prime (x^\prime)} {\partial x^{\prime \mu}} $$
$$ \ \ \ \ \ = \frac {\partial \phi (x)} {\partial x^{\prime \mu}}=  \frac {\partial x^\nu} {\partial x^{\prime\mu}}\frac {\partial \phi (x)} {\partial x^ \nu}\eqno(7)$$
 Clearly, the transformation given by eq.(7)  is different from the one in eq.(6). This motivates one to introduce another kind of vector field called covariant vectors. 
 
 A covariant vector field $V_\mu (x)$ is defined to be an object such that under   $x^\gamma \rightarrow x^{\prime \gamma}$, 
$$ V_\mu (x) \rightarrow V^\prime _\mu (x^\prime) = \frac {\partial x^\nu} {\partial x^{\prime\mu}}V_\nu (x)\eqno(8)$$

To summarize, while the transformation property and the directional nature of  infinitesimal vector $dx^\mu $  leads to the notion of contravariant vectors, similar considerations concerning the partial derivative $\frac {\partial } {\partial x^\mu}$ entails the concept of covariant vectors. In 3+1 dimensional space-time, vector fields have 4 components corresponding to $\mu = 0,1,2,3$. In pictorial terms, contravariant vectors are like arrows while the covariant vectors are like normal vectors to surfaces.

We can now wrap up the above considerations to  arrive at a generalization -  tensor fields of arbitrary ranks. A tensor field $V^{\mu_1 \mu_2 ...\mu_{n-1} \mu_n} _{\ \ \ \ \nu_1 \nu_2 ...\nu_{m-1} \nu_m} (x) $ of  rank $n+m$  is an entity such that under the coordinate transformation $x^\alpha \rightarrow x^{\prime \alpha}$, 
$$V^{\mu_1 \mu_2 ...\mu_{n-1} \mu_n}_{\ \ 
\ \ \nu_1 \nu_2 ...\nu_{m-1} \nu_m} (x) \rightarrow V^{\prime\mu_1 \mu_2 ...\mu_{n-1} \mu_n} _{\ \ \ \ \nu_1 \nu_2 ...\nu_{m-1} \nu_m} (x^\prime)$$
where,
$$ V^{\prime\mu_1 \mu_2 ...\mu_{n-1} \mu_n} _{\ \ \ \ \nu_1 \nu_2 ...\nu_{m-1} \nu_m} (x^\prime) = \frac {\partial x^{\prime\mu_1}} {\partial x^{\alpha_1}}  \frac {\partial x^{\prime\mu_2}} {\partial x^{\alpha_2}}... \frac {\partial x^{\prime\mu_n}} {\partial x^{\alpha_n}} \frac {\partial x^{\beta_1}} {\partial x^{\prime\nu_1}}  \frac {\partial x^{\beta_2}} {\partial x^{\prime\nu_2}} ...  \frac {\partial x^{\beta_m}} {\partial x^{\prime\nu_m}}V^{\alpha_1 \alpha_2 ...\alpha_{n-1} \alpha_n}_{\ \ \ \ \beta_1 \beta_2 ...\beta_{m-1} \beta_m} (x)\eqno(9)$$
We should note that in the above equation the arguments $x^\alpha $ and $x^{\prime \alpha}$ of $V$ and $V^\prime $, respectively, are the coordinates of the same event, as emphasized in the first paragraph of this section. In other words, transformation of tensors are completely local because of which tensors of identical ranks can be added and subtracted.

 An important result that follows from eq.(9) is that if a tensor vanishes at an event in one coordinate system, it is identically zero at that event in all coordinate systems. 
   
A fundamental entity in GTR that describes space-time geometry is the space-time dependent metric tensor $g_{\mu \nu}(x^\alpha)$,  which  determines the invariant proper distance $ds$
 between any two nearby events with coordinates $x^\mu$ and $x^\mu + dx^\mu $,

 $$ds^2 = g_{\mu \nu}(x) dx^\mu dx^\nu \eqno(10)$$
 Here, $x^\mu $, $\mu, \nu$=0,1,2,3, now represents a general curvilinear coordinate, specifying the location of an event. The metric $g_{\mu \nu} (x) $ is a generalization of  $\eta_{\mu \nu}$,
 the Minkowski metric tensor.
 
 Is $g_{\mu \nu} (x)$ a tensor field? If $ds^2$ is invariant under general coordinate transformation, then we can readily prove that $g_{\mu \nu} $ is a covariant tensor of rank 2. This is because, under $x^\alpha \rightarrow x^{\prime \alpha}$, we have, 
$$ds^2 = g_{\mu \nu} (x) dx^\mu dx^\nu \rightarrow ds^2 = g^\prime _{\alpha \beta} (x^\prime) dx^{\prime \alpha} dx^{\prime \beta}$$
$$= g^\prime _{\alpha \beta} (x^\prime) \bigg ( \frac {\partial  x^{\prime \alpha}} {\partial x^\mu} dx^\mu \bigg )\bigg ( \frac {\partial  x^{\prime \beta}} {\partial x^\nu} dx^\nu \bigg )$$ 
$$=  \frac {\partial  x^{\prime \alpha}} {\partial x^\mu}   \frac {\partial  x^{\prime \beta}} {\partial x^\nu} g^\prime _{\alpha \beta} (x^\prime) dx^\mu dx^\nu \eqno(11)$$
Comparing eqs.(10) and (11) as well as using the fact that $dx^\mu$ is an arbitrary infinitesimal separation, we get the result,
 $$g_{\mu \nu} (x)= \frac {\partial  x^{\prime \alpha}} {\partial x^\mu}   \frac {\partial  x^{\prime \beta}} {\partial x^\nu} g^\prime _{\alpha \beta} (x^\prime) \eqno(12)$$
implying that $ g_{\mu \nu} $ is a covariant tensor of second rank. This result readily connects with equivalence principle in the following manner.

  If the space-time geometry was not curved, one could choose a coordinate system such 
that everywhere the metric tensor is just the Minkowski metric tensor. But GTR states that energy and momentum associated with matter warp the space-time geometry, entailing that in general it is
not possible to choose inertial coordinates everywhere so that the metric is globally Minkowskian.

 However, according to the principle of equivalence, by choosing an appropriate coordinate system,
 even in an
arbitrarily curved space-time, the metric tensor can be made to take the form of $\eta_{\mu \nu}$ in a sufficiently small space-time region (physically, this corresponds to choosing a sufficiently
small freely falling frame). This is precisely what eq.(12) entails. One can choose a new set of coordinates $\xi^\alpha $ such that in a small region,
$$\eta_{\mu \nu} = \frac {\partial  x^{\prime \alpha}} {\partial \xi^\mu}   \frac {\partial  x^{\prime \beta}} {\partial  \xi^\nu} g^\prime _{\alpha \beta} (x^\prime) \eqno(13)$$

 For the dynamics of bodies moving in pure gravity, the notion of gravitational mass becomes  superfluous in GTR since particle  trajectories  are  geodesics of  space-time geometry determined from the line-element,
 $$ ds^2= g_{\mu \nu} dx^\mu dx^\nu \ $$
 Hence, it is not surprising that the world lines of freely falling test particles are independent of their inertial masses.

One can also define a contravariant metric tensor $g^{\mu \nu} (x) $ by demanding that,
$$ g^{\mu \alpha} (x) g_{\alpha \nu} (x) = \delta ^\mu _\nu \eqno(14)$$
If one considers $g_{\mu \nu} $ to be a $4 \times 4$ matrix, then eq.(14) implies that the contravariant metric tensor $ g^{\mu \nu}$ can be viewed as the inverse of the corresponding matrix.

Both $g_{\mu \nu} (x)$ and $g^{\mu \nu} (x)$ are symmetric tensor fields, i.e. $g_{\mu \nu} = g_{\nu \mu} $ and $g^{\mu \nu} = g^{\nu \mu} $ everywhere in the space-time manifold.

Employing $g^{\mu \nu} (x)$, one can raise indices of a covariant tensor, just like  lowering the indices of a  contravariant tensor field can be achieved by using $g_{\mu \nu} $. Therefore,
$$V^\mu (x) =  g^{\mu \nu} (x) V_\nu (x) \eqno(15)$$
is a contravariant vector field corresponding to the covariant vector field $V_\mu (x)$. While,
$$W_\mu (x) =  g_{\mu \nu} (x) W^\nu (x) \eqno(16)$$
is a covariant vector field corresponding to the contravariant vector field $W^\mu (x)$. Hence, raising and lowering of indices can be done freely for any tensor field of any rank by making suitable use of the metric tensors.

So far we have done some amount of tensor algebra. Let us now take up some tensor calculus. Suppose $A^\mu (x)$ is a contravariant vector field. Is $\frac {\partial A^\mu (x)} {\partial  x^\nu}$ a second rank tensor? Under a general coordinate transformation $x^\gamma \rightarrow x^{\prime \gamma}$,
$$\frac {\partial A^\mu (x)} {\partial  x^\nu} \rightarrow \frac {\partial A^{\prime \mu} (x^\prime)} {\partial  x^{\prime \nu}}=\frac {\partial  x^{\alpha}} {\partial x{\prime ^\nu}} \frac {\partial } {\partial x{^\alpha}} \bigg ( \frac {\partial x^{\prime\mu}} {\partial x^\beta} A^\beta (x) \bigg )  $$
$$=\frac {\partial  x^{\alpha}} {\partial x{\prime ^\nu}}   \frac {\partial x^{\prime\mu}} {\partial x^\beta} \frac {\partial A^\beta (x)} {\partial x{^\alpha}}+ \frac {\partial  x^{\alpha}} {\partial x{\prime ^\nu}}  \frac {\partial^2 x^{\prime\mu}} {\partial x^\alpha\ \partial x^\beta}\ A^\beta (x)\eqno(17)$$

It is obvious from eq.(17) that because of the second term in its right hand side,$\frac {\partial A^\mu (x)} {\partial  x^\nu}$ does not transform like a tensor. What is the remedy?

This is where the concept of covariant derivative comes in. We introduce a new  mathematical object $\Gamma^\mu_{\alpha \beta} (x^\lambda)$ referred to as Christoffel symbol (and also as affine connection and Levi Civita connection), and define the covariant derivative of $A^\mu (x)$ as follows,
$$A^\mu ;\nu = A^\mu ,\nu + \Gamma^\mu_{\nu \alpha} A^\alpha \eqno(18)$$
where,
$$A^\mu ,\nu \equiv    \frac {\partial A^\mu (x)} {\partial  x^\nu}\eqno(19)$$

{\bf (From now on partial derivative w.r.t. $x^\alpha $ will be denoted by $ , \alpha $)}

Using eq.(17), it can be easily shown that $A^\mu ;\nu $, defined by eq.(18), transforms as a tensor of rank 1+1 provided the  Christoffel symbol  $\Gamma^\mu_{\alpha \beta} (x^\lambda)$ transforms under a general coordinate transformation as,
 $$\Gamma^\mu_{\alpha \beta} (x^\lambda)\  \rightarrow \ \Gamma^{\prime \ \mu}_{\alpha \beta} (x^{\prime \lambda}) = \frac {\partial x^{\prime \mu}} {\partial x^\nu} \frac {\partial x^\sigma} {\partial x^{\prime \alpha}} \frac {\partial x^\gamma} {\partial x^{\prime \beta}} \  \Gamma^\nu_{\sigma \gamma} (x^\lambda) + \ \frac {\partial x^{\prime \mu}} {\partial x^\nu} \frac {\partial^2 x^\nu} {\partial x^{\prime \alpha} \partial x^{\prime \beta} }\eqno(20)$$
 
Clearly, because of the second term in the right hand side of eq.(20), Christoffel symbol is not a tensor. One can use precisely this feature to choose local Minkowski coordinates $\xi^\mu$ to make $\Gamma^\mu_{\alpha \beta}$ vanish at a space-time point. This dovetails nicely with equivalence principle, since we know that in a freely falling frame, the metric is $\eta_{\mu \nu} $ in a small neighborhood so that it has vanishing derivatives at a point.

Now, since for any arbitrary scalar field $\phi (x) $, $ \frac {\partial \phi (x)} {\partial x^\mu} $ is already a covariant vector field (see eq.(7)), the covariant derivative of any scalar field is its usual partial derivative, 
$$\phi ;\nu = \phi , \nu \eqno(21)$$

Now, if $V^\mu (x) $  and $U_\mu (x) $ are any two contravariant and covariant vector fields, respectively, $V^\mu (x)U_\mu (x) $ is a scalar field (see Prob.1(a)). Hence, its covariant derivative according to eqs. (18) and (21) is given by,
$$(V^\mu (x)U_\mu (x)); \nu = V^\mu;\nu  U_\mu + V^\mu U_\mu; \nu = (V^\mu (x)U_\mu (x)), \nu $$
$$  = V^\mu , \nu U_\mu + V^\mu U_\mu, \nu $$
$$ = (V^\mu ,\nu + \Gamma^\mu_{\nu \alpha} V^\alpha) U_\mu + V^\mu U_\mu; \nu \eqno(22) $$
From eq.(22) it ensues,
$$U_\mu; \nu = U_\mu ,\nu - \Gamma^\alpha_{\nu \mu} U_\alpha \eqno(23)$$
as $V^\mu (x) $ is an arbitrary contravariant vector field. This procedure can be deployed to obtain covariant derivatives of any tensor field of arbitrary rank. 

Therefore, as particular examples,
$$T_{\mu \nu}; \alpha  = T_{\mu \nu}, \alpha  - \Gamma^\beta_{\alpha \mu} T_{\beta \nu} - \Gamma^\beta_{\alpha \nu} T_{\mu \beta} \eqno(24)$$
and,
$$A^{\mu \nu}; \alpha  = A^{\mu \nu}, \alpha   + \Gamma^\mu_{\alpha \beta} A^{\beta \nu} +  \Gamma^\nu_{\alpha \beta} A^{\mu \beta} \eqno(25)$$

In 3-dimensional Euclidean geometry, the line-element $dl^2= dx^2+dy^2 + dz^2 $ has
the same form whether you shift the Cartesian coordinate system by any constant vector or rotate the coordinate system about any axis by any constant angle. The line-element given by eq.(2) is
 similarly invariant under  Lorentz transformations as well as constant space-time translations. According to the equivalence principle, whatever is the  gravity around, in a locally inertial
 frame (i.e. freely falling frame), the line-element is given by eq.(2) and non-gravitational laws of physics take the same form as in special relativity. 
  But, what is the connection between this feature of gravitation and geometry?

 Consider a  generally curved two-dimensional surface (e.g. the surface of, say, a pear). No matter how greatly the surface is curved, one can always choose a tiny 
enough patch on it, such that it is
 sufficiently flat for Euclidean geometry to hold good over it. As one increases the size of the patch, the curvature of the pear's surface  becomes apparent.  This is
 so similar to the main characteristic of gravity  that  we discussed in the
 preceding paragraph. The small patch on the pear over which the line-element is Euclidean ($dl^2= dx^2 + dy^2 $) is analogous to the local inertial frame in the case of 4-dimensional space-time
 where the line-element is described by eq.(2).

\section{Christoffel Symbol, Curvature Tensor and the Einstein equations}
Now, from eqs.(15) and (16), in order that,
$$V^\mu ; \alpha  =  g^{\mu \nu} ; \alpha  V_\nu +  g^{\mu \nu} V_\nu ; \alpha = g^{\mu \nu} V_\nu ; \alpha $$
and,
$$W_\mu ; \alpha =  g_{\mu \nu} ; \alpha  W^\nu  + g_{\mu \nu} W^\nu  ; \alpha =  g_{\mu \nu} W^\nu  ; \alpha $$
we require,
$$ g^{\mu \nu} ; \alpha  = 0 = g_{\mu \nu} ; \alpha \eqno(26)$$
Making use of eq.(26), we have,
$$g_{\mu \alpha} ; \nu + g_{\alpha \nu} ; \mu - g_{\mu \nu} ; \alpha =0 \eqno(27)$$
From the eqs.(24) and (27), it can be easily proved that,
$$\Gamma^\mu _{\alpha \beta} =  \frac {1} {2} g^{\mu \lambda} (g_{\alpha \lambda}, \beta + g_{\beta \lambda}, \alpha  - g_{\alpha \beta}, \lambda )\eqno(28)$$
displaying an important fact that Christoffel symbol is related to metric tensor and its derivatives.

Because of eq.(28), one can show that the trajectory (i.e. the worldline) $x^\mu (\lambda) $, where $\lambda $ is an affine parameter characterizing the worldline, that extremizes the proper length (invariant under general coordinate transformations),
$$S \equiv  \int {ds} = \int {\sqrt{g_{\mu \nu} dx^\mu dx^\nu}}= \int {\sqrt{g_{\mu \nu} \frac {dx^\mu} {d\lambda} \frac {dx^\nu} {d\lambda}} d\lambda}\eqno(29)$$ 
satisfies the geodesic equation,
$$\frac {d^2 x^\alpha} {d\lambda^2} + \Gamma^\alpha _{\mu \nu} \frac {dx^\mu} {d\lambda} \frac {dx^\nu} {d\lambda}=0 \ .\eqno(30)$$
 
In weak fields (small departure from Minkowski space-time), one can choose quasi-Minkowskian coordinates so that,
$$ g_{\mu \nu} = \eta_{\mu \nu} + h_{\mu \nu} \eqno(31)$$
with,
$$ \vert h_{\mu \nu} \vert  \ll 1 \eqno(32)$$
for $\mu, \nu $=0,1,...,3. 

For static and weak gravitational fields where test particles move with speeds much less than $c$, one must have eq.(30) reduce to Newtonian gravitational dynamics, where a particle with spatial coordinate $x^i, \ i=1,2,3$ satisfies,
$$\frac {d^2 x^i} {dt^2} = - \frac {\partial  \phi_N (\vec {r})} {\partial x^i}\eqno(33)$$
where $\phi_N (\vec {r} )$ is the Newtonian gravitational potential at $\vec {r}$.

Eq.(30) indeed leads to eq.(33) in the weak and static field approximation provided,
$$ g_{00} = 1 + h_{00} \approx 1 +\frac {2 \phi_N (\vec {r} )} {c^2} \eqno(34)$$
$$ g_{0i} \approx 0 \ , \ \ \ g_{ij} \approx - \delta _{ij}\eqno(35)$$
so that,
$$ \Gamma ^i_{00} \approx \frac {1} {c^2} \frac {\partial  \phi_N} {\partial x^i} \eqno(36)$$

We know from STR that the time elapsed in a clock (comoving with an observer O') cruising with uniform velocity with respect to an  inertial observer O is given by,
$$\tau =  \frac{1} {c} \int {\sqrt{\eta_{\mu \nu} dx^\mu dx^\nu}} = \sqrt { 1 - v^2/c^2} \ \ t \ ,\eqno(37)$$
where $x^\mu $, $v$ and $t$  are the space-time coordinates of O', speed of O' and time as measured by the inertial observer O, respectively. This is   called the proper time that elapses in the frame of  O', and is invariant under Lorentz transformations. The time dilation result ensues from eq.(37). 

What can we say about the proper time elapsed for a test particle as it moves along an arbitrary  worldline in a curved space-time? Let the worldline in a space-time manifold whose geometry is described by  metric $ g_{\mu \nu} (x^\alpha) $ be described by  $x^\mu (\lambda) $ from $\lambda_1$ to $\lambda_2$, $\lambda $ being an  affine parameter characterizing the worldline. 

Since one understands good clocks and good measuring rods in the framework of STR, the way to measure  proper time $\tau $ elapsed in an arbitrarily accelerating clock in arbitrary gravity is clearly by adding the infinitesmal time intervals elapsed in local inertial frames that lie along the trajectory of the clock at different instants of time and that co-move with the clock at those instants of time (for comoving clocks $v=0$ so that the proper time is just the time elapsed in these clocks as seen from eq.(37)). But by virtue of eqs.(11) and (13), this sum   is just,
$$\tau=\frac{1} {c} \int^2_1 {\sqrt{g_{\mu \nu} \frac {dx^\mu} {d\lambda} \frac {dx^\nu} {d\lambda}} d\lambda} \eqno(38)$$
Another way of stating the above argument is     that the infinitesimal proper time interval between two time-like separated close by events is $ds/c$, and since time is additive,  the total proper time elapsed is simply the integral  given by eq.(38).
 
We can apply the above result to determine the proper time elapsed in a clock at rest in a weak and static gravity. Using eqs.(34) and (35) in eq.(38) for a clock at rest ($dx^i=0$) at a point A, one obtains the proper  time elapsed to be  given by,
$$\tau_A \approx (1 + \frac {\phi_N (A)} {c^2} )\ \  t  \ ,\eqno(39)$$
 $t$ being the proper time elapsed in the frame of a static inertial observer at infinity where $h_{00}=0$. From this one concludes that not only time runs slow in attractive gravitational fields but also radiation emitted from regions with stronger and attractive gravitational potentials get redshifted as they move out to weaker    gravity regions.

To summarize, in order to connect  LIFs at different space-time points, and to express physical laws in terms of arbitrary coordinates in reference frames of  size  as large   as one wishes, one needs the language of tensor calculus  so that one  acquires an affine connection $\Gamma^\mu _{\alpha \beta}$ derivable from the metric tensor $g_{\mu \nu} $ and its derivatives. Although, this affine connection (or, Christoffel symbol) vanishes at a point in a LIF, its derivative does not.

This brings us to the Riemann curvature tensor $R^\mu _{\ \nu \alpha \beta} $ which represents how curved is the space-time geometry, and is constructed out of Christoffel symbol and its derivatives in the following way,
$$R^\mu _{\ \nu \alpha \beta} = \Gamma^\mu _{\nu \beta} , \alpha  -  \Gamma^\mu _{\nu \alpha} , \beta   +  \Gamma^\mu _{\sigma  \alpha}   \Gamma^\sigma _{\nu \beta}  - \Gamma^\mu _{\sigma  \beta}   \Gamma^\sigma _{\nu \alpha} \eqno(40)$$
From eq.(40), it is obvious that,
$$R^\mu _{\ \nu \alpha \beta} = - R^\mu _{\ \nu \beta \alpha}$$
One can obtain a symmetric  second rank tensor called the Ricci tensor from the Riemann tensor,
$$R_{\nu \beta} = R^\mu _{\ \nu \mu \beta} = R_{\beta \nu } \eqno(41) $$ 
The Ricci scalar is simply,
$$R = g^{\mu \nu} R_{\mu \nu} \eqno(42)$$ 
From eq.(40), one can easily prove that if,
$$R_{\mu \nu \alpha \beta} = g_{\mu \lambda} R^\lambda _{\ \nu \alpha \beta}$$ then,
$$R_{\mu \nu \alpha \beta}= - R_{\mu \nu \beta \alpha} =  - R_{  \nu  \mu\alpha \beta} =  R_{\alpha \beta \mu \nu }\eqno(43) $$

To summarize, Christoffel symbol is like gravitational field since it involves derivatives of the metric. However, because it is not a tensor (see eq.(20)), it can also represent pseudo-gravity. For instance,
consider a  hypothetical case in which there is no gravity anywhere, so that one may choose global Minkowskian coordinates in order that the metric tensor is simply $\eta_{\mu \nu}$ everywhere  in this coordinate system. In such a situation, for an observer O' accelerating with respect to an inertial observer O,  one can easily show that the Christoffel symbol for O' is non-zero and, in fact,  corresponds to a fictitious force in the accelerating frame that mimics gravity. 
 
  The true gravitation, represented by the Riemann curvature tensor, is the tidal gravity. 
Although in a local inertial frame, gravity  disappears, tidal gravitational force does not. Since the Christoffel symbol is not a tensor,  in a LIF it is zero at a point, while the Riemann tensor in general  is nonzero. This reminds us of the acceleration due to gravity vanishing while the tidal gravitational force being nonzero, in a LIF.

For instance, earth is freely falling towards the sun because of latter's pull. But we do not feel sun's
gravity since the freely falling earth constitutes a local inertial frame. However, as sun's gravity is non-uniform, portions of earth closer to the sun feel a greater tug
 than those  located farther. This differential pull is the source of tidal force which causes the commonly observed ocean tides. In GTR, the tidal acceleration is due to the fourth rank
  Riemann tensor that is
constructed out of the metric and its first as well as second derivatives. Therefore, the ocean tides owe their existence  to the non-zero Riemann tensor describing the 
  curvature of space-time geometry around the sun (as well as the moon).
  
  Therefore,   true gravity  represented by the tidal gravitational field is related to the Riemann curvature tensor $R^\mu _{\nu \alpha \beta} $, a fourth rank tensor  constructed out of the connection $\Gamma^\mu _{\alpha \beta}$ and its derivatives. In mathematics, $R^\mu _{\nu \alpha \beta} $  determines whether the geometry is flat or curved.  This, in a sense,  completes the identification of gravity with geometry. While in the gauge theory framework, $\Gamma^\mu _{\alpha \beta}$ is analogous to   gauge potential with $R^\mu _{\nu \alpha \beta} $ as the corresponding gauge covariant field strength. 
 
In later lectures, we will see that the dynamics of space-time geometry is determined by the Einstein equations,
 $$R_{\mu \nu} - \frac {1} {2} R g_{\mu \nu} = \frac {8 \pi G} {c^4} T_{\mu \nu}\eqno(44)$$
 where the Ricci tensor and Ricci scalar are  $R_{\mu \nu} \equiv R^\alpha  _{\mu \alpha \nu} $ and $R= g^{\mu \nu} R_{\mu \nu}$, respectively. $T_{\mu \nu} $ is the matter energy-momentum tensor whose various  components represent the flux of energy and momentum carried by matter in appropriate directions. When the gravity is weak and static, eq.(33) reduces to Newton's gravity,
 $$\nabla^2 \phi = 4 \pi G \rho \eqno(45)$$
 for a non-relativistic source with mass density $\rho $ and negligible pressure. The Newtonian gravitational potential $\phi $ is identified with the geometrical entity $(g_{00} - 1) c^2/2$.

  GTR tells us that matter distorts the space-time from an Minkowskian geometry to a non-Minkowskian one, and test bodies just move along
 straightest possible paths in such a curved space-time. As to, how the matter warps the space-time geometry, is given by the so called Einstein equations
 which relate tensors created out of the metric and the Riemann tensor to the matter energy-momentum tensor multiplied by a combination of Newton's constant G and light speed c.
 
 Einstein equations possess a pristine beauty, with space-time geometry on one side, and the energy and momentum of matter on the other.
 When the geometrical curvature of space-time is small and the motion within the source  is slow enough, GTR leads automatically to Newton's laws of gravitation.
 
\section{Gravitational Radiation} 
 GTR as a theory of gravitation gained  immediate acceptance among the physics community as soon as its prediction of bending of light was actually seen
 during the solar eclipse of 1919. Of course, GTR had already correctly explained the anomalous precession of the perihelion of Mercury. Since GTR is based on special relativity, gravitational perturbations too propagate as gravitational waves (or, undulations
in space-time geometry) with finite speed c. Later, indirect evidence for  gravitational waves  predicted by 
Einstein was corroborated with the discovery of slowly inspiralling Hulse-Taylor binary pulsar (PSR 1913 + 16). 

%Very far away from a source whose energy and momentum distributions are changing asymmetrically, if the ensuing perturbation in the space-time geometry, represented by $h_{\mu \nu} $,  is   sufficiently weak, one can choose a quasi-Minkowskian coordinate system and express the metric tensor as,
%$$ g_{\mu \nu} \approx \eta_{\mu \nu} + h_{\mu \nu} (\vec r, t) \ ,\eqno(46)$$
%with the perturbation or the gravitational wave amplitude satisfying,
%$$ \vert h_{\mu \nu} (\vec r, t) \vert  \ll 1 \eqno(47)$$
%at large distances from the source.

 %The gravitational wave (GW) amplitude $h_{\mu \nu} (\vec r, t)$ is determined by the second   time derivative of the mass quadrupole moment of the source that is  undergoing changes in its matter distribution. Since  the metric tensor governs the proper distance (or equivalently, proper time) by virtue of eq.(29), the proper  distance between  two test particles  will undulate when a GW is incident on  them. By measuring the relative separation between the test particles as a function of time, one can gain information about $h_{\mu \nu}$.
 
Very far away from a source whose energy and momentum distributions are changing asymmetrically, if the ensuing perturbation in the space-time geometry, represented by $h_{\mu \nu} $,  is   sufficiently weak, one can choose a quasi-Minkowskian coordinate system and express the metric tensor as,
$$ g_{\mu \nu} \approx \eta_{\mu \nu} + h_{\mu \nu} (\vec r, t) \ ,\eqno(46)$$
with the perturbation or the gravitational wave amplitude satisfying,
$$ \vert h_{\mu \nu} (\vec r, t) \vert  \ll 1 \eqno(47)$$
at large distances from the source.

 The gravitational wave (GW) amplitude $h_{\mu \nu} (\vec r, t)$ is determined by the second   time derivative of the mass quadrupole moment of the source that is  undergoing changes in its matter distribution. Physical effects of  GWs on test particles are best understood in the transverse, traceless (TT) gauge. In this gauge, only the space-space components of the GW amplitude are non-zero along with the conditions of vanishing trace and components orthogonal to the direction of propagation. Hence, if the    GW is propagating in the z-direction, it can have only $h_{11}= - h_{22}$ and $h_{12}= h_{21}$ as the non-zero components (so that they are manifestly traceless and orthogonal to the z-direction).  
 
%In general, for a source with non-relativistic internal motion and mass density $\rho(t, \vec r)$, the GW amplitude in the TT-gauge as seen by an observer at time t and position $\vec r $ ($r \gg $ source size) from the source  is given by [1],
%$$h^{ij} (t, \vec r) = \frac{2 G} {c^4 r} \frac {d^2 \not{I}^{ij} (t - r/c)}{dt^2} \eqno(48a)$$  
%where the reduced mass quadrupole moment is defined as,
%$$\not{I}^{ij}(t) \equiv I^{ij}(t) - \frac {1} {3} \delta^{ij} I^k_k (t)\eqno(48b)$$
%with the mass quadrupole moment being,
%$$I^{ij} (t) \equiv \int {\rho(t, \vec r) x^i x^j d^3r} \ \ .\eqno(48c)$$
%In eq.(48a), causality is ensured because of the retarded time $t - r/c$ appearing in the RHS. 
 
%Do GWs transport energy? Now, only entities that have energy can possibly be perceived or measured, since exchange of energy between an object and the sensors is crucial for its detection. Even in quantum theory, two subsystems can influence each other only via an interaction Hamiltonian. Feynman and Hermann Bondi had used the following thought experiment to demonstrate that GWs carry energy [1]: Consider two loose metal rings around a rod that is held in a horizontal position. If a GW passes by,  the rings  will move and oscillate  with respect to the rod (elasticity of the rod will prevent appreciable change in its length because of the incident GW).  Hence, the rings and the rod will get heated up because of friction. This energy certainly has to be at the expense of the energy carried by the GW.
 
 Do GWs transport energy? Now, only entities that have energy can possibly be perceived or measured, since exchange of energy between an object and the sensors is crucial for its detection. Even in quantum theory, two subsystems can influence each other only via an interaction Hamiltonian. So, in order   for  a measurement device to determine an  eigenvalue corresponding to a quantum observable of a system, there has to be an exchange of energy between the system and the apparatus through a suitable interaction Hamiltonian.
 
  Returning to GWs,   Feynman and Hermann Bondi had used the following thought experiment to demonstrate that GWs transport  energy [1]: Consider two loose metal rings around a rod that is held in a horizontal position. If a GW passes by,  the rings  will move and oscillate  with respect to the rod (elasticity of the rod that gives the latter rigidity will prevent appreciable change in its length because of the incident GW).  Hence, the rings and the rod will get heated up because of friction. This energy certainly has to be at the expense of the energy carried by the GW. In fact, the observed slow decrease in the orbital period of the two  neutron stars in the Hulse-Taylor binary pulsar PSR 1913+16 demonstrates unequivocally that  the loss of  binary system energy  is due to the radiated GWs that carry energy away, agreeing  extremely well with the  prediction ensuing from GTR [2].

In the TT-gauge, for a source with non-relativistic internal motion and mass density $\rho(t, \vec r)$, the GW amplitude  as seen by an observer at time t and position $\vec r $ ($r \gg $ source size) from the source  is given by [3],
$$h^{ij} (t, \vec r) = \frac{2 G} {c^4 r} \frac {d^2 (\not{I}^{ij} (t - r/c))}{dt^2} \eqno(48a)$$  
where the reduced mass quadrupole moment is defined as,
$$\not{I}^{ij}(t) \equiv I^{ij}(t) - \frac {1} {3} \delta^{ij} I^k_k (t)\eqno(48b)$$
with the mass quadrupole moment being,
$$I^{ij} (t) \equiv \int {\rho(t, \vec r) x^i x^j d^3r} \ \ .\eqno(48c)$$
In eq.(48a), causality is ensured because of the retarded time $t - r/c$ appearing in the RHS.

The quantity $\frac {d^2 \not{I}^{ij} (t - r/c)}{dt^2}$ is, in some sense, a measure of asymmetric motion of    matter in the GW source, representing approximately the non-symmetric portion of the source's kinetic energy. Hence, from eq.(48a), one can write down a  formula to make back-of-the-envelope estimate of the GW  magnitude, 
$$ h \approx \frac {4 G E_{nonsym}} {c^4 \ r} \ \ .\eqno(49)$$

The energy-momentum pseudo-tensor corresponding to GWs is given by [3],  
$$T_{\mu \nu}=\frac {c^4} {32\pi G} \bigg < h_{jk, \mu} h^{jk}_{,\nu} \bigg > \eqno(50)$$
where $h_{jk}$ is the GW amplitude in the TT-gauge and that, $<...>$ represents averaging over many wavelengths (Raising and lowering of indices of GW amplitude are done using Minkowski metric tensor.).

 The  difficulty of constructing a proper energy-momentum tensor for the gravitational degrees of freedom is related to the fact that $g_{\mu \nu}\rightarrow \eta_{\mu \nu}$ and $\Gamma^\mu_{\alpha \beta}\rightarrow 0$ (at a point)  in a local inertial frame (due to Eintein's equivalence principle) so that any second rank tensor built out of $g_{\mu \nu}$ and $\Gamma^\mu_{\alpha \beta}$ will be simply zero at a point P in a local inertial frame. But this point P is arbitrary since one can choose a local inertial frame anywhere in the entire space-time. Hence, it implies that such a tensor is identically zero every where. If one includes first derivatives of $\Gamma^\mu_{\alpha \beta}$, then  symmetric tensors like $R_{\mu \nu}$ or $G_{\mu \nu}$ can certainly be constructed. But $R_{\mu \nu}$ and $G_{\mu \nu}$ vanish where there is no matter by virtue of Einstein equations (eq.(44)), belying their representing energy and momentum flux of GWs propagating through vacuum. Therefore, it is common practice to employ the Landau-Lifshitz energy-momentum pseudo-tensor, that contains only first derivatives of the metric tensor, to study energy and momentum associated with gravitational degrees of freedom [4].

Since $T_{00}$ is the energy density, the GW energy flux is given by,
$$F_{GW}= c T_{00}= \frac {c^5} {32\pi G} \bigg < h_{jk, 0} h^{jk}_{,0} \bigg >=
\frac {c^3} {32\pi G} \bigg < \dot h_{jk} \dot h^{jk} \bigg >  \eqno(51)$$
where  $\dot h_{jk} \equiv \frac {\partial h_{jk}} {\partial t}$. Making use of eq.(48a) in eq.(51) one obtains,
$$F_{GW}= \frac {1} {4 \pi r^2} \bigg (\frac {G} {2 c^5}  \bigg < \dddot{\not{I}}_{jk}\dddot{\not{I}}^{jk} \bigg > \bigg )\eqno(52)$$
When the background space-time is nearly flat (so that $4 \pi r^2 \cong$ surface area of a sphere of radius $r$) and the emission of GWs is isotropic, it is evident from eq.(52) that the GW luminosity is given by,
$$L_{GW}\cong 4 \pi r^2\ F_{GW}=\frac {G} {2 c^5}  \bigg < \dddot{\not{I}}_{jk}\dddot{\not{I}}^{jk} \bigg > \ \ .\eqno(53)$$
Now, if mass distribution in a GW source changes over a typical time scale of $\sim T_{GW} = 2 \pi/\omega=1/f$, then from eq.(53) one can pen down a back-of-the-envelope expression to estimate $L_{GW}$, 
$$L_{GW} \sim \frac {2 G} { c^5} \omega^2 E^2_{nonsym} \approx 2 \times 10^{50} \ \bigg (\frac {E_{nonsym}} {10^{51}\ \mbox{erg\ s}^{-1}} \bigg )^2 \bigg ( \frac {f} {1\ \mbox{kHz}} \bigg )^2 \ \ \mbox{erg\ s}^{-1}\ \ .\eqno(54) $$
\subsection{GW Luminosity, $c^5/G$, Planck Scales and Hawking Radiation}
It is very interesting to observe that the physical dimensions of both $\frac { c^5}  {G}=3.6 \times 10^{59} \mbox{\ erg \ s}^{-1}$ and $\dddot{\not{I}}_{jk}$ are identically $\frac {M L^2} {T^3} = [\mbox{Luminosity}]$. Does this  indicate  something deep about gravitation?
%$\bullet$  An interesting feature concerning physical dimensions:
%\vskip 1 em
%$(\frac {G} { c^5})^{-1}$ and $\frac{\partial^3 \not{I}_{jk}} {\partial t^3} $ are identically $\frac {M L^2} {T^3} = [\mbox{luminosity}]$
%\vskip 1 em

As an amusing exercise related to the above coincidence, one may consider the gravitational collapse of a compact astrophysical object like a supra-massive neutron star of mass $M$ and initial radius $R=\alpha_1 R_s$, where $\alpha_1$ is a number just in excess of unity  and $ R_s\equiv 2 G M/c^2$ is the Schwarzschild radius. If the collapse to a black hole is non-spherical, with  asymmetric kinetic energy $E_{nonsym}= \alpha_2 M c^2$ (where $\alpha_2$ is a number less than but of order unity), and takes place in dynamical time scale $\sqrt{R^3/GM} \sim T_{GW}$  then, according to eq.(54), 
$$L_{GW}\sim \frac{\alpha^2_2} {4 \alpha^2_1}\ \frac { c^5} {G}= 9 \times 10^{58} \ \frac{\alpha^2_2} { \alpha^2_1}\  \mbox{erg \ s}^{-1}\eqno(55a)$$
and  $\dddot{\not{I}}_{jk} \sim   E_{nonsym}/T_{GW}\sim \frac{\alpha_2} {2 \sqrt{2} \alpha_1}\ \frac { c^5} {G}$,  is also roughly of the same order as $L_{GW}$. In this case, curiously enough, neither $L_{GW}$ of eq.(55a) nor  $ E_{nonsym}/T_{GW}$ depend on the mass of the  compact, collapsing object. Hence, formation of primordial black holes, either due to collision of bubble walls or rapid collapse of false vacuum pockets in the early universe [5,6], could generate GW luminosity comparable to $c^5/G$. Similarly, compact supra-massive neutron stars (likely  progenitors of fast radio bursts) undergoing near free fall, as they collapse very rapidly to form black holes,  might also lead to such high GW luminosity [7].

A further connection to $c^5/G$ emerges, if one considers Planck energy, $E_{Pl} \equiv m_{Pl} c^2 \equiv \sqrt{c^5 \hbar / G}$ and the Planck time, $t_{Pl} \equiv \sqrt{\hbar G/ c^5}$ so that  one may define Planck luminosity, $L_{Pl}$ [8],
$$L_{Pl}\equiv \frac {E_{Pl}}{t_{Pl}}=\frac{c^5} {G} \ . \eqno(55b)$$
Eq.(55b) indicates three interesting features - (i) close to the time of big bang when quantum gravity effects were dominant, quantum fluctuations could have generated GWs with luminosity $\sim c^5/G$, (ii) Planck luminosity does not have the quantum imprint $\hbar $ and, hence, (iii) for an astrophysical source to generate such a large luminosity, if its non-symmetric kinetic energy $E_{nonsym}$ scales as $ \sim \alpha_3 E_{Pl}$, the time scale over which its mass quadrupole moment changes substantially must also scale as $\sim \alpha_3 t_{Pl}$, $\alpha_3$ being a very large positive number. 

Above considerations  suggest that $ c^5/G $ may represent the maximum  possible  GW luminosity (see also [8]). If it embodies a fundamental upper limit for $L_{GW}$  then it leads to a lower limit for the time scale  $T_{GW}$ (or equivalently, an upper limit on the characteristic frequency $f$) over which matter in a GW source gets redistributed. This follows from,
$$L_{GW} \approx \frac {E_{nonsym}} {T_{GW}} = \frac {\alpha_2 M c^2} {T_{GW}} \leq \frac{c^5}{G}$$ 
$$\Rightarrow T_{GW} \geq \frac {\alpha_2 G M} {c^3}=1.5 \times 10^{-5} \ \alpha_2 \ \bigg (\frac {M} {3 \ M_\odot} \bigg )\ \mbox{s} $$
$$\Rightarrow f \leq \frac {c^3}{\alpha_2 G M}\ \cong  67 \ \alpha^{-1}_2 \ \bigg (\frac {M} {3 \ M_\odot} \bigg )^{-1} \ \mbox{kHz} \ \ .\eqno(55c)$$ 

Although  GW luminosity may  approach $c^5/G $ in situations involving  collapse to form primordial black holes or implosion of supra-massive neutron stars, the total energy radiated in the form of GWs will be limited by $E_{nonsym}=\alpha_2 M c^2$. One may also compare Planck luminosity  with the luminosity associated with  the evaporation of black holes that was predicted by Stephen Hawking in 1974 [5,6].  Existence of Hawking temperature, $T_H = c^3 \hbar /8 \pi G M k_B$,  for a black hole of mass $M$ implies that the event horizon acts like a  black body,  emitting radiation with a flux, $F_H=\sigma T^4_H$,  so that the corresponding luminosity is given by,
$$L_H=F_H\times 4 \pi R^2_s=\frac {1} {15360 \pi} \bigg (\frac {m_{Pl}} {M} \bigg )^2 \frac{c^5}{G} \ .\eqno(55d)$$
From eq.(55d) it is evident that even for a Planck scale primordial black hole, the luminosity corresponding to Hawking evaporation is four orders of magnitude below the Planck luminosity.

Since most of the GW sources that are going to be detected  are likely to be extra-galactic (e.g. even the first directly detected GWs from GW150914 originated at a cosmological redshift of $z\approx 0.09$), it is instructive to generalize the expression given in eq.(52) for sources lying at cosmological distances. To do so, we will make use of GW luminosity [9].

Imagine an extra-galactic  source placed at a Robertson-Walker radial coordinate $r$. If this object radiates GWs with a luminosity density $\mathcal{L}_{GW}(t, \nu)$ at cosmic time $t$ then energy radiated in the  time interval $(t, t+ \Delta t)$ and frequency band $(\nu, \nu + \Delta \nu)$ is $\mathcal{L}_{GW}(t) \Delta t \Delta \nu$. Luminosity $L_{GW} (t)$ is related to the luminosity density $\mathcal{L}_{GW}(t, \nu)$ as,
$$L_{GW} (t)=\int^\infty_0 {\mathcal{L}_{GW}(t, \nu) d\nu}\eqno(56)$$
Hence, the number of gravitons emitted in the time interval $(t, t+ \Delta t)$ and frequency range $(\nu, \nu + \Delta \nu)$is given by,

$$\Delta N (t, \nu) = \frac {\mathcal{L}_{GW}(t, \nu) \Delta t \Delta \nu} {h \nu}\eqno(57)$$

Assuming that GWs are radiated isotropically from the source, the radiated energy and the number of gravitons will be spread over an area $4 \pi a^2 (t_0) r^2 $ by the time it reaches an observer ($r=0$) at present times over the time interval $(t_0, t_0+ \Delta t_0)$. Due to the expansion of the universe,  both frequency $\nu$ and $\Delta t$ will be cosmologically redshifted to $\nu_0 = \nu \frac {a(t)} {a(t_0)}= \frac {\nu} {1+z}$ and $\Delta t_0 = \Delta t \frac {a(t_0)} {a(t)} = (1+z) \Delta t $, respectively. (Here we have used the Robertson-Walker coordinates with $a(t)$ being the expansion scale factor and redshift $z= \frac {a(t_0)} {a(t)} -1 $.)

The number of gravitons received at $r=0$ per unit area  in the time interval $(t_0, t_0+ \Delta t_0)$ and  frequency band $(\nu_0, \nu_0 + \Delta \nu_0)$ then is given by,
$$\Delta \mathcal{N} = \frac {\Delta N (t, \nu)} {4 \pi a^2 (t_0) r^2} = \frac {\mathcal{L}_{GW}(t, \nu) \Delta t \Delta \nu} {h \nu  \ 4 \pi a^2 (t_0) r^2}\eqno(58)$$
Therefore, GW energy received in the frequency range $(\nu_0, \nu_0 + \Delta \nu_0)$ per unit area per unit time is given by,
$$\mathcal{F}(t_0, \nu_0) \Delta \nu_0= \frac { h \nu_0 \Delta \mathcal{N}}{\Delta t_0} = \frac { h \nu_0 \mathcal{L}_{GW}(t, \nu) \Delta t \Delta \nu} {h \nu \ 4 \pi a^2 (t_0) r^2 \Delta t_0 } =  \frac {\mathcal{L}_{GW}(t, \nu)  \Delta \nu} { 4 \pi a^2 (t_0) r^2 (1+z)^2}\eqno(59)$$ 
Thus, from eqs.(56) and (59), the GW flux is simply,
$$F_{GW} (t_0) = \int^\infty_0 {\mathcal{F}(t_0, \nu_0) d\nu_0} = \frac {L_{GW}(t)} { 4 \pi a^2 (t_0) r^2 (1+z)^2}=\frac {L_{GW}(t)} { 4 \pi D^2_L(z)}\ \ ,\eqno(60)$$ 
 the required generalization of eq.(52), where $D_L(z) $ is the standard luminosity distance of Friedmann-Robertson-Walker cosmology.  

 %So, if a linearly polarized GW is propagating in the z-direction, it can have only $h_{11}= - h_{22}$ and $h_{12}= h_{21}$ as the non-zero components that are orthogonal to the direction of propagation. Consider now, for simplicity, the case of a near monochromatic GW for  which  $h_{11}= - h_{22}=h$ and $h_{12}= h_{21}=0$.
%Suppose one has two test particles scattered on the xy-plane. In order to monitor the change in the proper distance  between the test particles  because of the metric perturbation, one has to have the initial separation $L$  between these  particles to be much less than the  radius of curvature of space-time geometry associated with the incident GW, so that one can use either eq.(29) or the geodesic deviation equation to determine the change in the separation. 
\subsection{Measuring GW amplitude}
Physical effects of  GWs on test particles are best understood in the transverse, traceless gauge.
Since  the metric tensor governs the proper distance (or equivalently, proper time) by virtue of eq.(29), the proper  distance between  two test particles  will undulate when a GW is incident on  them. By measuring the relative separation between the test particles as a function of time, one can gain information about $h_{\mu \nu}$. Consider now, for simplicity, the case of a near monochromatic GW for  which  $h_{11}= - h_{22}=h$ and $h_{12}= h_{21}=0$. Suppose one has two test particles scattered on the xy-plane. In order to monitor the change in the proper distance  between the test particles  because of the metric perturbation, one has to have the initial separation $L$  between these  particles to be much less than the  radius of curvature of space-time geometry associated with the incident GW, so that one can use either eq.(29) or the geodesic deviation equation to determine the change in the separation.

For a GW, the radius of curvature of space-time geometry is of the order of its wavelength. Hence, the above condition then is,
 $$ L \ll c (\omega /2\pi)^{-1} \ \eqno(61)$$
 where $\omega $ is the angular frequency of the GW. If the two test particles lie on the x-axis then according to eq.(29), the proper distance $l(t)$ is given by,
 $$l(t)= \int {\sqrt{- g_{\mu \nu} dx^\mu dx^\nu}}= \int^L_0 {\sqrt{1 - h_{11}} \ dx}\cong \bigg (1 -  \frac {1} {2}\ h \bigg ) L \ ,\eqno(62)$$ 
 where one has used the condition given by eq.(47). The negative sign inside the square-root sign in eq.(62) occurs because the two test particles at the same instant of time are space-like separated.
 
  Therefore, the strain corresponding to the change in length is simply related to the GW amplitude in the following manner,
 $$\frac {\Delta L} {L} \equiv \frac {l(t) - L}{L} = - \frac {1} {2}\ h \ .\eqno(63)$$
 Eq.(62) tells us that as the GW passes by, the proper distance between the test particles changes with time. Therefore, if light is emitted from one test particle at time $t_1$ 
  towards the other, the arrival time $t_2$ of light at the other test particle is given by,
  $$t_2 \approx t_1 + l(t_1)/c = t_1 +  \bigg (1- \frac {1} {2}\ h \bigg ) L/c \ .\eqno(64)$$
 It is essentially this feature, represented by eq.(64), that is used in a laser interferometer to detect GW, since changes in the arrival times from the two arms of the interferometer correspond directly to the  changes in the phase difference that lead to GW amplitude measurements from the fringe shifts in the interference pattern (see [10] for a pedagogical introduction to this subject).

  In a time interval $T= 2 \pi/\omega$ (where $\omega$ is the angular frequency of the near monochromatic GW), the variation in the proper distance  $\Delta L$ between two test particles intercepting the GW can change by an amount of the order of $h L/2$ (eq.(63)), where $h$ and $L$ are the magnitude of GW amplitude and initial separation between the particles (provided $L$ is much much less than the wavelength of the GW, which is $\sim 2\pi c/\omega$). An interesting question to ask is: Can the change $\Delta L$ happen so rapidly that $\Delta L/T > c$? 
Now, $\Delta L/T= h L/2T = h L \omega/4 \pi$, which is much less than $h c$ because of eq.(61).
Since, $h$ is much much less than unity (eq.(47)), $\Delta L/T$ is much much less than $c$. Therefore, the rate of change of separation between the test particles can never exceed the speed of light.

\section{The first direct measurement of GW amplitude}

The two LIGOs,  laser interferometric  GW detectors of USA,   independently achieved  direct detection of GWs from  a black hole binary merger event on September 14, 2015,  with a time delay of about 6.9 ms [11]. This  GW  source is at a redshift $z\approx  0.09$,  corresponding to a luminosity distance of about 410 Mpc. The  GW150914 binary system consisted of two coalescing  black holes of mass  $ 29 \pm 4\ M_\odot $ and $ 36^{\ + 5}_{\ - 4} \ M_\odot $ that eventually collided with each other to settle down into a bigger rotating  black hole of mass $ M_f = 62 \pm 4 \   M_\odot $.  The spin angular momentum of the final black hole is estimated to be $ 0.67 ^{\ + 0.05}_{\ - 0.07} \  \frac {G M^2_f} {c} $.  The mass deficit of about 3 $M_\odot $ was carried away by GWs. Significantly, this event not only corroborates GW results that ensue from GTR but is also consistent with the prediction    of quasi-normal mode emission of GWs from a perturbed black hole [12].

Therefore, it is pertinent  to ask whether  GW150914 is  consistent with black hole thermodynamics (BHT) too. A theorem proved by Hawking  several decades back implies, in this case, that  area of the event horizon of the final black hole must be larger than the sum of the event horizon areas of the binary components. The radius of the event horizon of a black hole of mass $M$ and spin angular momentum $L$ is given by (see, for example,[13]),
$$R_{EH}= \frac {G} {c^2} \bigg (M + \sqrt{M^2 - (L \ c/G\ M)^2} \bigg ) \eqno(65a)$$
while the area of the event horizon is given by (see, for instance,[14]),
$$A= 8 \pi \bigg (\frac {G} {c^2} \bigg)^2 M \bigg (M + \sqrt{M^2 - (L\ c/G\ M)^2} \bigg )\eqno(65b) $$

 There is considerable uncertainty about the spin angular momenta of the two initial black holes of GW150914.  In our analysis, we may take them to be Schwarzschild black holes (i.e. $L=0$) so that we may start with the maximum event horizon area. Then, from eq.(65b) with $L=0$, the  initial  total area of the event horizons is  given by, 
$$A_i = 16 \pi (G/c^2)^2 [M^2_1 + M^2_2] \eqno(66)$$
where $M_1 = 29 \ M_\odot $ and  $M_2 = 36 \ M_\odot \ .$  
The black hole formed after the merger has a mass $M_f = 62 \ M_\odot $ and spin angular momentum $L_f = 0.67 \ \frac {G M^2_f} {c} $.
Hence, the final area of the event horizon according to eq.(65b) is given by,
$$A_f =  8 \pi (G/c^2)^2 M_f \bigg (M_f + \sqrt{M^2_f - (L_f \ c/G\ M_f)^2} \bigg )\eqno(67)$$
Therefore, from eqs.(66) and (67), the ratio of final area to the initial is,
$$\frac {A_f}{A_i} = \frac { M^2_f \bigg (1+ \sqrt{1 - (L_f\ c/G \ M^2_f)^2} \bigg )}{2 \bigg (M^2_1 + M^2_2 \bigg )} = 1.57 \  \eqno(68)$$
where one has used,
$$L_f\ c/G\ M^2_f = 0.67 \ .$$

Thus, even after overestimating the initial area by assuming that the initial black holes were Schwarzschild black holes, eq.(68) demonstrates that the parameters deduced from the event GW150914 are consistent with Hawking's area theorem, which states that in any classical physical process,  the event horizon area of the final black hole must be larger than the sum of the event horizon areas of the initial black holes involved in the process (see, for example, [14]).

 One may go one step ahead by including the estimated errors in the parameters, and check  if Hawking's black hole area theorem is violated in the worst case scenario by 
 considering  $M_1 = 33 \ M_\odot $,  $M_2 = 41 \ M_\odot \ $, $M_f = 58 \ M_\odot $ and $L_f = 0.72 \ \frac {G M^2_f} {c}$ (consistent with the errors quoted in [11]).
It turns out then,
$$\bigg (\frac {A_f}{A_i} \bigg )_{\mbox{min}} = 1.03 \ ,\eqno(69)$$
which is still in agreement with the area theorem (I have made a simplifying assumption in the above analysis that the errors in the parameters are mutually independent).

On the other hand, if one considers the case where $M_1 = 25 \ M_\odot $,  $M_2 = 32 \ M_\odot \ $, $M_f = 66 \ M_\odot $ and $L_f = 0.6 \ \frac {G M^2_f} {c}$ (again, being consistent with the errors reported in [11]), one finds that,
$$\bigg (\frac {A_f}{A_i} \bigg )_{\mbox{max}} = 2.38 \ .\eqno(70)$$
The right hand side of  eq.(70) can, of course,  be larger if the initial black holes are  Kerr black holes. One may combine eqs.(69) and (70) to express the ratio as,
$$ \frac {A_f}{A_i} = 1.57 ^{\ + 0.81}_{\ - 0.54} \ ,$$
that suggests  GW150914 upholding the classical BHT. 

LIGOs, subsequently, have detected GWs from three more black hole binary mergers, with the latest been observed by VIRGO, an European GW detector,  too. In a straight forward exercise, analogous to the preceding analysis, one can easily verify that the parameters deduced from the three later GW detections  are in agreement with Hawking's black hole area  theorem.   
\section{GWs from stellar outflows}
In a very recent paper, it has been shown that a carbon-rich red giant star, V Hydrae, is linked with ejection of fast moving  ($V=$200 km/s to 250 km/s) plasma blobs as heavy as planet Mars [15]. The carbon star is at a distance of about $r=500$ pc from us [16]. The paper argues that ejection of these knotty outflows originate from an accretion disc around an unseen companion  of V Hydrae that is moving in a highly  eccentric orbit around the common centre of mass with an orbital period of 8.5 yrs. The plasma `cannonballs' are shot out when the companion of mass $M_c < M_p$ passes close to the stellar envelope of the primary (of mass $M_p=$1 to 2  $M_\odot$) at periastron [15].    

It may be an instructive exercise to calculate  the GW amplitude expected from such a source using  the back-of-the-envelope estimate discussed in section VI. For the V Hydrae case, the mass $m_b$ of the outflowing blobs are found to be $\approx (0.7 - 1.0) \times 10^{27}$ gm [15]. Therefore,
$$ E_{nonsym} = \frac {1} {2} m_b V^2 \approx 3 \times 10^{41} \bigg ( \frac {m_b}{10^{27}\ \mbox{gm}} \bigg )  \bigg ( \frac {V}{250\ \mbox{km/s}} \bigg )^2\ \mbox{erg}.\eqno(71)$$
Estimates of GW amplitude from sources with  relativistic jets   have been made in the past (e.g. [17] and [18]). In the case of V Hydrae, the plasma `bullets' shot out have non-relativistic speeds, and hence, one  uses eq.(49) to find  GW amplitude to be,
$$h \approx 7 \times 10^{-29} \bigg ( \frac {m_b}{10^{27}\ \mbox{gm}} \bigg )  \bigg ( \frac {V}{250\ \mbox{km/s}} \bigg )^2 \bigg ( \frac {r}{500\ \mbox{pc}} \bigg )^{-1} \eqno(72)$$
Since the blobs are ejected from the innermost radius $r_a$ of the accretion disc, typical time scale characterizing the acceleration of these blobs is given by,
$$\Delta t \sim \frac {2 \pi r_a} {V} \lesssim  1.75 \times 10^4 \bigg ( \frac {r_a} {R_\odot} \bigg ) \ \mbox{s}\ ,\eqno(73)$$
assuming that $r_a$ is of the order of  the size of the companion star (which is likely to be a main sequence star of solar or sub-solar mass [16]). Eq.(73) implies that the GW amplitude from these plasma knots, given by eq.(72),  would  have a characteristic frequency $\nu_c \sim 1/\Delta t \gtrsim  0.1 $ mHz.  According to the theoretical model proposed by Sahai et al., such outbursts from V Hydrae binary system are likely to occur  every 8.5 yrs, whenever the companion grazes past the stellar envelope of the primary [15]. Therefore, an ultra-sensitive space based  GW detector of future (10 to 11 orders of magnitude more sensitive than the proposed eLISA) is likely to pick up such signals periodically.

Currently, it is a moment to rejoice as the recent direct measurements of GWs have culminated  in this year's  Nobel Prize in physics going to Rainer Weiss, Barry C. Barish and Kip S. Thorne.  
\vskip 2.0 em
{\bf{References}}
\vskip 1.0 em
1. D. Kennefick, `Controversies in the History of  the Radiation Reaction problem in General Relativity' (2007)  (http://dafix.uark.edu/~danielk/History/eins.pdf)
\vskip 1.0 em
2.  T. Damour \& J. H. Taylor, ApJ, 366, 501 (1991)
\vskip 1.0 em
3. K. S. Thorne, in Three Hundred Years of Gravitation (Cambridge University Press, 1989)
\vskip 1.0 em
4. L. D. Landau and E. M. Lifshitz, Classical Theory of Fields (Pergamon Press, 1975)
\vskip 1.0 em
5. S. W. Hawking, Nature, 248, 30 (1974)
\vskip 1.0 em
6. N. Upadhyay, P. Das Gupta and R. P. Saxena, Phys. Rev. D, 60, 063513 (1999) 
\vskip 1.0 em
7. P. Das Gupta and N. Saini, to appear in J. Astrophys. Astron., arXiv:1709.00185 [astro-ph.HE]
\vskip 1.0 em
8. C. J. Hogan,  in Structure Formation in the Universe (Springer Netherlands, 2001)
\vskip 1.0 em
9. P. Das Gupta and J. V. Narlikar, Mon. Not. R. Astron. Soc., 264, 489-496 (1993)
\vskip 1.0 em
10. Peter R. Saulson, American Journal of Physics, 65, 501 (1997)
\vskip 1.0 em
11. B. P. Abbott et al., Phys. Rev. Lett., 116, 241102 (2016)
\vskip 1.0 em
12. C. V. Vishveshwara, Nature (London), 227, 936 (1970)
\vskip 1.0 em
13. James B. Hartle,  Gravity: An Introduction to Einstein's General Relativity (Addison-Wesley, 2003)
\vskip 1.0 em
14. T. Jacobson (1996), http://www.physics.umd.edu/grt/taj/776b/lectures.pdf
\vskip 1.0 em  
15. R. Sahai, S. Scibelli \& M. R. Morris,  arXiv 1605.06728v1 [astro-ph.SR]
\vskip 1.0 em
16. Knapp, G. R., Dobrovolsky, S. I., Ivezić, Z., Young, K., Crosas, M.,
Mattei, J. A., \& Rupen, M. P., A \& A, 351, 97 (1999)
\vskip 1.0 em
17. Ehud B. Segalis and Amos Ori, Phys. Rev. D 64, 064018 (2001) 
\vskip 1.0 em
18. Ofek Birnholtz \& Tsvi Piran, Phys.Rev. D87,  123007 (2013)
\vskip 4.0 em 
\section*{Practice Problems}
\vskip 2.0 em
1 If $V^\mu (x) $ and $W_\mu (x) $ are two vector fields, then show that:

(a) $V^\mu W_\mu $ is a scalar field.

(b) $V^\mu W_\nu $ is a tensor field of rank 1+1

(c)$ g^{\nu \alpha} V^\mu W_\alpha$ is a tensor field of rank 2.

2 (a) A tensor $A^{\mu \nu}$ is found to be anti-symmetric in a particular coordinate system $\lbrace x^\mu \rbrace $. Prove that 
$A^{\mu \nu}= - \ A^{\nu \mu}$ in all coordinate systems. 

(b) It is given that a zero-rest mass particle is moving along the world line $x^\mu (\lambda) $ in 3+1-dimensions, where $\lambda $ is an affine parameter. Show that the tangent vector $u^\mu = \frac {dx^\mu} {d\lambda} $ satisfies the equation,
$$ g_{\mu \nu}u^\mu u^\nu = 0\ .$$ 

(c) (i) If $A^\mu _\nu (x)$ is a second rank tensor field of (1+1) type, then show that,
$$A^\mu _ \nu ; \alpha  = A^\mu _ \nu, \alpha   + \Gamma^\mu_{\alpha \beta} A^\beta_ \nu  - \Gamma^\beta _{\alpha \nu} A^\mu _\beta $$

(ii) Show that $\delta^\mu _ \nu ; \alpha = 0$

 3 (a) Consider the 2-dimensional manifold constituted by the surface of a sphere of radius $r=1$. 

(i) Choosing a convenient coordinate system, obtain $g_{\mu \nu}$, $g^{\mu \nu}$ and $\Gamma^\mu_{\alpha \beta} $. 

(ii) A vector $v^\mu$ starting out with components $v^1 = A$ and $v^2 = B$ from the point P: $(\theta, \phi) $ $ = (\pi/2, 0)$  is parallel transported from P to  $(\pi/2, \pi/2)$ first, then to $(\alpha, \pi/2)$, and thereafter to $(\alpha, 0)$, and finally back to P. Here, $0 \leq \alpha < \pi/2$. Find the components of  $v^\mu$ when it returns to P. Does your result change when $\alpha \rightarrow 0$? 

(b) For a vector field $ V^\mu (x^\gamma) $, prove that,
$$  V^\mu ;\alpha ; \beta \ - \  V^\mu ; \beta ;\alpha = - \ R^\mu _{\ \nu \alpha \beta} V^\nu $$ 

(c) If $\psi (x^\lambda)$ is a scalar field then prove that,
$$ g^{\mu \nu} \psi ;\mu ; \nu = \frac {1} {\sqrt{-g}}\ \frac {\partial} {\partial x^\nu}\bigg (\sqrt{-g}\ g^{\mu \nu} \frac {\partial \psi} {\partial x^\mu} \bigg ) $$

(d) For weak and static gravity, show that clocks tick at a slower rate in regions of stronger gravity. 

4 (a) The Einstein-Hilbert action is given by,
$$A_G = -\ \frac {c^3}{16 \pi G} \int {R \sqrt{- g} \ d^4 x} $$

Prove that under an infinitesimal variation 
$g^{\mu \nu} \rightarrow g^{\mu \nu} + \delta g^{\mu \nu}$, the variation in $A_G$ is given by,
$$\delta A_G = -\ \frac {c^3}{16 \pi G} \int {(R_{\mu \nu} - 1/2 \ g_{\mu \nu} R ) \delta g^{\mu \nu} \sqrt{- g} \ d^4 x} $$
provided $ \delta g^{\mu \nu}(x^\alpha) $ vanishes at infinity. 

(b) If $\delta g_{\mu \nu} $ is an infinitesimal variation of the metric tensor $g_{\mu \nu}$ and $T^{\mu \nu}$ is the energy-momentum tensor, prove that
$$T^{\mu \nu}\delta g_{\mu \nu}= - \ T_{\mu \nu}\delta g^{\mu \nu}$$ 

(c) If $\bar {G}^{\mu \nu} \equiv R^{\mu \nu} - 1/2 \ g^{\mu \nu} R + \Lambda g^{\mu \nu} $ then prove that $\bar {G}^{\mu \nu} ;\nu = 0$. Assume that $\Lambda $ is a constant.

5 (a) Show that a light ray emitted radially
 outward from $ r < R_s = 2\ G\ M/c^2 $ can never cross the event horizon of a  Schwarzschild blackhole of mass $M$.

 (b) For a Schwarzschild blackhole of mass $M$, show that a test particle of non-zero rest mass cannot have constant $r$ trajectories when it is inside the event horizon (i.e. $r  < R_s \equiv 2\ G\ M/c^2)$).

 5 (a) For a particular space-time, $\xi ^\mu (x^\alpha)$ is given to be a Killing vector field. Consider a test particle falling freely in this space-time along a geodesic $x^\mu (\lambda) $,   where $\lambda $ is an affine parameter. Show that $u^\mu \xi_\mu $ is a constant of motion, given that  $u^\mu \equiv \frac {dx^\mu} {d\lambda} $.

(b) Suppose the metric tensor $ g_{\mu \nu}$ is independent of the particular coordinate $x^\sigma$ for a fixed value of  $\sigma $ so that $\frac {\partial  g_{\mu \nu}} {\partial x^\sigma} = 0 $. Then, show that $\xi^\mu = \delta^\mu_{\ \sigma} $ satisfies the Killing equation,
$$\xi_{\mu \ ; \nu} \ + \ \xi_ {\nu \ ; \mu} =0 $$ 

(b) Given a spherically symmetric white dwarf of mass $2\times 10^{33} $ gm and radius 6000 km, find the maximum energy that can be extracted by lowering a test particle of mass $10^6$ gm very very slowly towards the white dwarf, by first obtaining an expression for the conserved energy of the test particle at rest in this space-time.

6 (a) Show that  a test particle of rest mass $m$ falling freely due to  gravity of a spherically symmetric body of mass $M \gg m$ moves along a geodesic $r(\phi)$ that satisfies the differential equation,
$$ \mbox{(i)}\ \ \ \ \ \ \  \frac {d^2 u} {d\phi^2} \ + \ u = \frac {R_s} {2 l^2} \ + \ \frac {3 R_s u^2} {2} $$
where $l\equiv \frac {L_z} {m c} $, $u(\phi) \equiv 1 /r(\phi)$, $R_s \equiv 2\ G\ M/c^2$ and $L_z$ is the angular momentum of the test particle. 

(ii) Under what approximation is,
$$r(\phi) \cong \frac { a (1-e^2)} {1+ e \cos {\phi}} $$
a solution of the differential equation in (i)? [$a$ and $e$ are positive constants] 

 (b) In the case of static and weak gravity, $g_{\mu \nu} \approx \eta_{\mu \nu} \ + \ h_{\mu \nu}$ with $\vert h_{\mu \nu} \vert \ll 1$ and $\partial h_{\mu \nu}/ \partial t = 0$. If $ h_{0 0} = 2 \phi_N/ c^2 $ show that Newton's gravity,
 $$\nabla^2   \phi_N = 4 \pi G \rho $$
 follows from the Einstein equation,
 $$R_{\mu \nu} = \frac {8 \pi G} {c^4} [T_{\mu \nu} - 1/2 \ g_{\mu \nu} T] $$
 where $\phi_N$ and $\rho $ are the Newtonian gravitational potential and mass density, respectively.

(c) Consider a blackhole of mass $M = 2 \times 10^{34}$ gm and a test body of mass $m=10^{-4}\ M$ orbiting around the blackhole  in the $\theta  = \pi/2 $ plane, with a closest radial approach $r_{min} = 10^{12}$ cm when it has the maximum speed  $v_{max} = 10^{-3}\ c$.  The test body follows, to a good approximation, the geodesic 
 $$ r(\phi) \cong \frac { 2 l^2/ R_s} {1+ e \ \psi (\phi)} $$
 where 
 $$\psi (\phi)\equiv \cos {\phi} + \frac {3 R^2_s} {4 l^2} \ \phi \ \sin {\phi} \ ,$$
 
 $ l \equiv \frac {L_z} {m c} $, $u(\phi) \equiv 1 /r(\phi)$, $R_s \equiv 2\ G\ M/c^2$, $e$ is the eccentricity and $L_z$ is the angular momentum of the test body. The orbital period $P$ is given to be $10^6$ seconds. Estimate the rate of precession of the point of closest approach. Does the result depend on $m$?

\end{document}